# Diverse outcomes of planet formation and composition around low-mass stars and brown dwarfs

Y. Miguel,[1]★ A. Cridland,[1] C. W. Ormel,[2] J. J. Fortney[3] and S. Ida[4]

[1]*Leiden Observatory, University of Leiden, Niels Bohrweg 2, NL-2333CA Leiden, the Netherlands*
[2]*Department of Astronomy, Tsinghua University, Beijing 100084, China*
[3]*Department of Astronomy and Astrophysics, University of California, Santa Cruz, CA 95064, USA*
[4]*Earth-Life Science Institute, Tokyo Institute of Technology, Meguro-ku, Tokyo 152-8550, Japan*

## ABSTRACT

The detection of Earth-sized exoplanets around low-mass stars – in stars such as Proxima Centauri and TRAPPIST-1 – provide an exceptional chance to improve our understanding of the formation of planets around M stars and brown dwarfs. We explore the formation of such planets with a population synthesis code based on a planetesimal-driven model previously used to study the formation of the Jovian satellites. Because the discs have low mass and the stars are cool, the formation is an inefficient process that happens at short periods, generating compact planetary systems. Planets can be trapped in resonances and we follow the evolution of the planets after the gas has dissipated and they undergo orbit crossings and possible mergers. We find that formation of planets above Mars mass and in the planetesimal accretion scenario, is only possible around stars with masses $M_\star \geq 0.07 M_{\rm sun}$ and discs of $M_{\rm disc} \geq 10^{-2} M_{\rm sun}$. We find that planets above Earth-mass form around stars with masses larger than $0.15\,M_{\rm sun}$, while planets larger than $5\,M_\oplus$ do not form in our model, even not under the most optimal conditions (massive disc), showing that planets such as GJ 3512b form with another, more efficient mechanism. Our results show that the majority of planets form with a significant water fraction; that most of our synthetic planetary systems have 1, 2, or 3 planets, but those with 4, 5, 6, and 7 planets are also common, confirming that compact planetary systems with many planets should be a relatively common outcome of planet formation around small stars.

## 1 INTRODUCTION

While theoretical calculations predicted that small rocky planets are the most common outcome of planet formation (e.g. Miguel, Guilera & Brunini 2011b), observing such planets present enormous technical challenges and only in recent years we started this new era of rocky exoplanet's detection and characterization. The discovery of Earth-sized planets orbiting low mass stars such as the planets in the TRAPPIST-1 system (Gillon et al. 2016), Proxima Centauri b (Anglada-Escudé et al. 2016), and the recent discovery of two planets around the star Teegarden (Zechmeister et al. 2019), present a unique opportunity to study the properties of such worlds. In addition, statistical studies show that small, rocky planets are ubiquitous around low mass stars (Bonfils et al. 2013; Kopparapu 2013; Dressing & Charbonneau 2015), and since these are the most common stars in the Galaxy (Henry et al. 2006; Winters et al. 2015), we are bound to discover many more of these planets in the close future.

This paper aims to explore the planet formation process and learn the most common trends and properties of the population of exoplanets hosted by M stars and brown dwarfs. Our results are applicable to systems such as the TRAPPIST-1 and similar systems, as well as the expected outcome of dedicated surveys such as the SPECULOOS project,[1] and space missions like TESS (Ricker et al. 2015) and PLATO (Rauer et al. 2014).

### 1.1 Similarities between Solar system satellites and formation of planets around small stars

The process of planet formation around low mass stars is poorly known. Because this process depends strongly on the central star, the pathway of planets around M stars and brown dwarfs might be more similar to the formation of regular satellites than the one around solar-type stars (Chiang & Laughlin 2013; Kane, Hinkel & Raymond 2013). Some evidence of this can be seen in the observed population of planets around small stars. This population is characterized by compact systems with planets orbiting in a few weeks at most, their orbits are co-planar (Fang & Margot 2012; Tremaine & Dong 2012) and most of them are rocky planets with very little gas[2] (Rogers 2015; Dittmann et al. 2017; Dorn et al. 2018; Grimm et al. 2018; Unterborn et al. 2018). In addition, Fig. 1 shows that all the exoplanet systems with masses measured with radial velocities and orbiting small stars, have a mass-ratio between $10^{-5}$ and $3 \times 10^{-4}$, comparable to the ones of the satellites around the giant planets. We note that the big difference of one order of magnitude with the Solar system is because most of the mass in the Solar system planets comes from the gas within the gas giants, and taking into account only the rocks might lead to a ratio of around $\simeq 2 \times 10^{-4}$, reinforcing the idea that one of the main differences is the possibility of having gas accretion. Nevertheless, we note that there are some uncertainties in the mass of heavy elements contained in Jupiter and the other giants and this number – marked with dashed lines in Fig. 1 – should be taken as an approximation (e.g. Fortney & Nettelmann 2010; Baraffe et al. 2014; Miguel, Guillot & Fayon 2016).

---

★ E-mail: ymiguel@strw.leidenuniv.nl

[1]https://www.speculoos.uliege.be/cms/

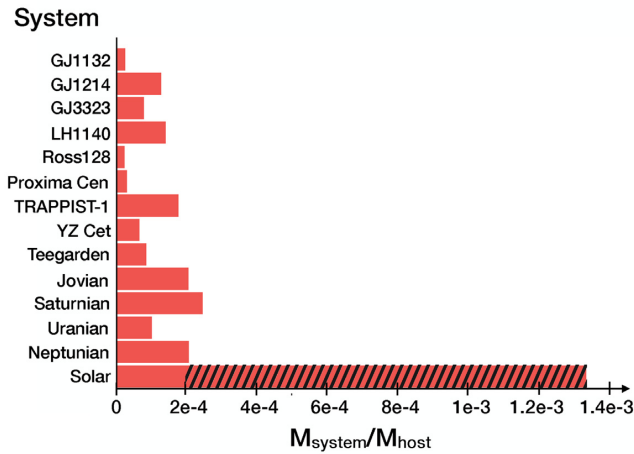

**Figure 1.** Mass ratio of the planets/satellites in different systems compared to that of the central star/planet. The dashed area marks the approximate mass of gas in the giant planets in the Solar system. Data by Berta-Thompson et al. (2015) for GJ1132, Charbonneau et al. (2009) for GJ1214, Astudillo-Defru et al. (2017) for GJ3323, Dittmann et al. (2017) and Ment et al. (2019) for LHS1140, Bonfils et al. (2018) for Ross128, Anglada-Escudé et al. (2016) for Proxima Centauri, Gillon et al. (2016) for TRAPPIST-1, Astudillo-Defru et al. (2017) for YZ Cet and Zechmeister et al. (2019) for Teegarden's star system.

Because they are fast and allow the exploration of many thousands of calculations, population synthesis simulations provide a unique way of exploring how the different formation scenarios impact and explain the statistical properties of the observed population of exoplanets (Benz et al. 2014). Previous population synthesis studies on the formation of planets around M stars and brown dwarfs using planetesimal accretion were adapted from models for higher stellar masses, and focused on the first millions of years until the gas dissipated in their discs (Payne & Lodato 2007; Alibert & Benz

2017). In this paper we explore the formation of such systems adapting a population synthesis model previously used to study the formation of Galilean satellites (Miguel & Ida 2016) and following the evolution of the system for $10^8$ yr, long after the gas dissipated in the disc.

We find that the main differences with the formation of planets around solar type stars are that the ones formed around small stars accrete no gas, and because the stars are cold, the regions where migration can be reversed (transition between the viscous heating and irradiation mechanisms, and the snow line) occur at small semimajor axis, which affects the distribution of close-in planets. To explore this, we include in this paper the possibility of having planetary traps, although we find that they do not have a huge impact in our calculations. In addition, the time-scale for solid accretion is inversely proportional to the solids surface density and since these are low mass discs, the time-scales became too large too quickly, and the formation can only happen close or inside the snow line (at around ∼0.2 AU). Therefore, the most massive initial protoplanets arise at the location of the snow line – which is the region with the largest concentration of solids – but because the snow line is located at a very small semimajor axis in these cold stars, the whole formation and evolution of planets in these systems occurs within a few AUs. This gives rise to compact systems with short period planets and makes the effect of drifting of planetesimals due to gas drag a crucial effect to consider in the formation of these systems. Also, both the satellites in our Solar system and also the TRAPPIST-1 planets are trapped in resonances. We also include in our calculations this important mechanism that affects the dynamical evolution and final architecture of these systems.

## 2 MODELLING PLANET FORMATION

We model the formation of planets from small embryos to evolved planets after an evolution of the system of $10^8$ yr. In our model we use standard planetesimal accretion, as a difference to other publications that explore the possibility of formation of the TRAPPIST-1 planets using pebble accretion (Ormel, Liu & Schoonenberg 2017; Schoonenberg et al. 2019). It is not our intention to have an extremely detailed physical model, but rather a simple model that considers all relevant physics in a semi-analytical way, coupled in a consistent manner while conserving the overall properties of the final population. In this section we describe our prescription for the formation of planetary systems, based in the approach by Miguel & Ida (2016).

### 2.1 Protoplanetary disc: initial structure and evolution

#### 2.1.1 Initial disc

A proper treatment of the protoplanetary disc structure and evolution is essential because it affects the growth and composition of the planets and also their dynamical evolution during the first $10^6$–$10^7$ yr. These discs are very complex systems and the ones around low mass stars are difficult to observe and poorly understood. In this paper we based our model on observations made with *Herschel* and ALMA (Joergens et al. 2013; Ricci et al. 2014; Liu et al. 2015; Daemgen et al. 2016; Testi et al. 2016; van der Plas et al. 2016). Future development in the measurements and physics of disc processes will help to improve our calculations.

The disc temperature profile is mostly determined by the heating source. In our model, the protoplanetary disc has two components:

---

[2]with the exception of GJ 1214b that has a large gas percentage (Berta et al. 2012; Rogers 2015).

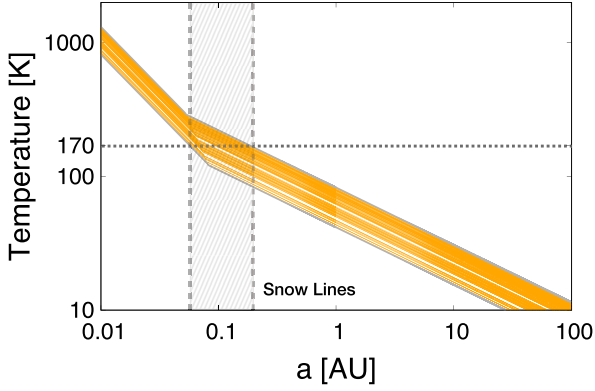

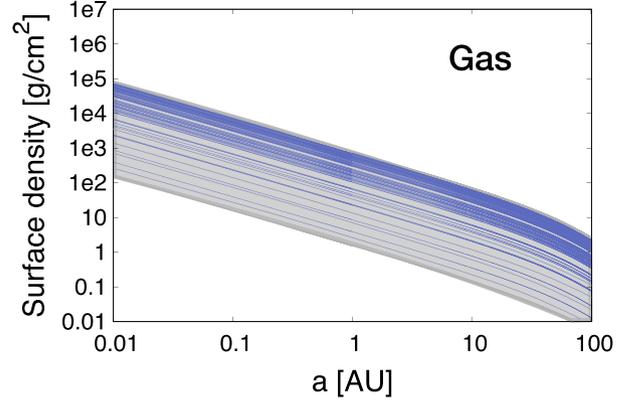

**Figure 2.** Temperature profile of 100 planetary systems (orange lines). In each system, the stellar mass was chosen randomly between 0.05 and 0.25 M$_{sun}$. The extreme profiles with the largest and smallest stellar masses are shown with grey thick lines. The temperature where water condenses is shown with the horizontal grey dotted line. The dashed vertical area covers all the possible snow lines.

**Figure 3.** Gas surface density versus semimajor axis of 100 planetary systems (blue lines). Similarly as in Fig. 2, these 100 systems are chosen randomly as examples of the discs used in the calculations. The full range of surface densities considered in this work is shown with the grey area.

an inner disc dominated by viscous heating, and an outer disc where the temperature is determined by the irradiation from the central star (Hueso & Guillot 2005; Oka, Nakamoto & Ida 2011; Ida, Guillot & Morbidelli 2016). We follow Ida et al. (2016) and adopt the following expression for the temperature of the disc in the viscous heating dominated region:

$$T_{vis} \simeq 127 \left(\frac{M_\star}{0.1\,M_\odot}\right)^{3/10} \left(\frac{\alpha}{10^{-3}}\right)^{-1/5} \left(\frac{\dot{M}_\star}{10^{-10}M_\odot\,yr^{-1}}\right)^{2/5}$$
$$\times \left(\frac{a}{0.1\,AU}\right)^{-9/10} K, \quad (1)$$

where $M_\star$ is the stellar mass, $\alpha$ is the parameter that characterizes the viscosity and $\dot{M}_\star$ is the accretion rate into the star, which has a typical value of $10^{-10}\,M_\odot\,yr^{-1}$ for small stars (Manara & Testi 2014). The irradiation temperature profile is given by:

$$T_{irr} \simeq 150 \left(\frac{M_\star}{0.1\,M_\odot}\right)^{3/7} \left(\frac{a}{0.1\,AU}\right)^{-3/7} K, \quad (2)$$

where we used the following relation between stellar mass and luminosity for small pre-main sequence stars (Ramirez & Kaltenegger 2014):

$$\frac{L_\star}{L_\odot} = \left(\frac{M_\star}{M_\odot}\right)^2. \quad (3)$$

The separation between the two regimes is given at:

$$a_{vis-irr} \simeq 0.067 \left(\frac{M_\star}{0.1\,M_\odot}\right)^{-3/11} \left(\frac{\alpha}{10^{-3}}\right)^{-14/33}$$
$$\times \left(\frac{\dot{M}_\star}{10^{-10}M_\odot\,yr^{-1}}\right)^{28/33} AU. \quad (4)$$

Fig. 2 shows the range of different temperature profiles considered in the calculations, where we see that $a_{vis-irr}$ is located very close to the star ($a_{vis-irr} < 0.1$ AU) for all the cases. We also see the change in the slope due to the two different heating regimes considered throughout the disc.

The snow line is the radius at which water condenses, increasing the surface density of solids in the outer parts of the disc. We calculate the position of the snow line using equation (2) and assuming that the water condensation temperature at the typical conditions in a protoplanetary disc is $T(a_{snow}) = 170$ K. Then we have:

$$a_{snow} \simeq 0.075 \left(\frac{M_\star}{0.1\,M_\odot}\right) AU. \quad (5)$$

For simplicity, we adopt steady accretion disc solution with constant $\alpha$ and parameterize the disc mid-plane gas surface density with a power law (Miguel, Guilera & Brunini 2011a; Miguel et al. 2011b):

$$\Sigma_g(a) = \Sigma_g^0(a)\left(\frac{a}{a_c}\right)^{-\gamma} e^{(\frac{a}{a_c})^{2-\gamma}}, \quad (6)$$

where we use $\gamma = 1$ and $a_c = 90$ AU in our calculations and $\Sigma_g^0$ is calculated from the total mass of the disc, $M_{disc}$:

$$M_{disc} = 2\pi \int_0^\infty a\,\Sigma_g(a)\,da. \quad (7)$$

Then:

$$\Sigma_g^0(a) = \frac{(2-\gamma)M_{disc}}{2\pi\,a_c^2}, \text{ with } \gamma < 2 \quad (8)$$

For the solids surface density we adopt a similar density profile at the disc mid-plane:

$$\Sigma_s(a) = \Sigma_s^0(a)\eta_{snow}\left(\frac{a}{a_c}\right)^{-\gamma} e^{(\frac{a}{a_c})^{2-\gamma}}, \quad (9)$$

where, according to estimations using solar photospheric and meteoritic abundances (Abe et al. 2000; Lodders 2003):

$$\eta_{snow} = \begin{cases} 2 & \text{if } a > a_{snow}, \\ 1 & \text{if } a < a_{snow} \end{cases} \quad (10)$$

and we adopt an initial gas to solids ratio of $\frac{\Sigma_g^0(a)}{\Sigma_s^0(a)} = 100$.

Figs 3 (gas surface density) and 4 (solids surface density) show density profiles for 100 systems chosen randomly from a linearly uniform distribution, which biases our results towards more massive discs. The full range of models considered is shown with the grey area. We see in Fig. 4 that the snow line is located extremely close to the central star ($a_{snow} < 0.2$ AU), which leads to the formation of very compact systems (see Section 3.1).

We assume that the disc is truncated at the magnetospheric cavity radius (Frank, King & Raine 1992; Ormel, Liu & Schoonenberg 2017):

$$a_{in} = 0.01 \left(\frac{B_\star}{180\,G}\right)^{4/7} \left(\frac{R_\star}{0.5\,R_\odot}\right)^{12/7} \left(\frac{M_\star}{0.1\,M_\odot}\right)^{-1/7}$$
$$\times \left(\frac{\dot{M}_\star}{10^{-10}\,M_\odot\,yr^{-1}}\right)^{-2/7} \text{AU}, \quad (11)$$

where $R_\star$ is the stellar radius and $B_\star$ is the strength of the magnetic field measured at the surface of the star. We assume that $B_\star \simeq 180\,G$, according to observations of brown dwarfs (Reiners, Basri & Christensen 2009).

*2.1.2 Evolution of the disc*

Because there are big uncertainties both in observations and theoretical studies of gas accretion into low mass stars, we adopt a simple model for the protoplanetary disc evolution where the depletion of the gas surface density follows an exponential decay (Ida & Lin 2004):

$$\Sigma_g(a) = \Sigma_g^0(a) e^{-\frac{t}{\tau_{disc}}} \quad (12)$$

with $\tau_{disc}$ the gas disc dissipation time-scale. Since the gaseous disc dissipates, the gas accretion rate into the star also decays exponentially:

$$\dot{M}_\star = \dot{M}_\star^i e^{-\frac{t}{\tau_{disc}}} \quad (13)$$

and this in turn affects $a_{vis-irr}$ and $a_{in}$, which move outwards with time.

The disc of solids is depleted locally due to planetary accretion:

$$\Sigma_s(a) = \Sigma_s^0(a) - \frac{M_p}{2\pi a\,10 R_H} \quad (14)$$

with $R_H$ the hill radius, $M_p$ the planetary mass, and $\Sigma_s^0(a)$ the initial solids surface density.

The disc of solids is also depleted globally because of the gas drag effect. Planetesimals orbit the star at a Keplerian speed and suffer a headwind caused by the gas that orbits at a slightly sub-Keplerian velocity. As a consequence, the planetesimals drift towards the star at a time-scale given by (Mosqueira & Estrada 2003; Miguel & Ida 2016):

$$\tau_{gas} \simeq 10^6 \left(\frac{\rho_s}{1\,g\,cm^{-3}}\right) \left(\frac{r_s}{1\,km}\right) \left(\frac{T}{100\,K}\right)^{-\frac{3}{2}} \left(\frac{\Sigma_g}{10^4\,g\,cm^{-2}}\right)^{-1}$$
$$\times \left(\frac{M_\star}{0.1\,M_\odot}\right) \text{yr}, \quad (15)$$

where $\rho_s$ is the planetesimals' typical density and $r_s$ their radius, assumed as 30 km in the calculations, close to the typical size distribution in the asteroid and Kuiper belt objects (Sheppard & Trujillo 2010) and that can also be reproduced by detailed simulations of streaming instability (Johansen et al. 2015; Abod et al. 2019).

**2.2 Planetary growth**

In the core accretion scenario with planetesimal accretion, the planetary embryos are embedded in the solid and gaseous disc and grow by accreting smaller planetesimals of some kilometres in size. In this scheme, the growth rate of the embryos is determined by the velocity dispersion of the swarm of planetesimals in its vicinity. In the dispersion-dominated regime, the encounters between the embryo and the planetesimals is well described by a 3D process and the planets accrete solids at a time-scale of (Ida & Lin 2004):

$$\tau_{acc} \simeq 2 \times 10^5 \left(\frac{a}{0.1\,AU}\right)^{1/2} \left(\frac{M_p}{1\,M_\oplus}\right)^{2/3} \left(\frac{M_\star}{0.1\,M_\odot}\right)^{-1/6}$$
$$\times \left(\frac{R_p}{1\,R_\oplus}\right)^{-1} \left(\frac{\Sigma_s}{100\,g\,cm^{-2}}\right)^{-1} \text{yr}. \quad (16)$$

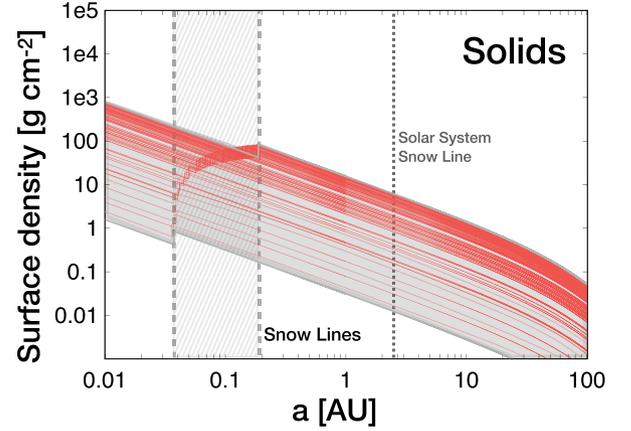

**Figure 4.** Solids surface density versus semimajor axis of 100 planetary systems (red lines). All possible snow lines are shown within the vertical dashed area. The Solar system snow line is shown as a comparison (dotted line).

Because of the small stellar and disc masses explored in this paper, the planetary embryos never reach the $\simeq 10\,M_\oplus$ necessary to start gas accretion (Ida & Lin 2004), therefore, we only form planets made by rocks or rocks and ices in our simulations.

**2.3 Migration**

The angular momentum exchange between the growing planetary embryos or protoplanets and the gaseous disc leads to an orbital migration of the protoplanets (Goldreich & Tremaine 1980). Here we have the combined effects of two torques: Lindblad torques due to spiral waves produced by gravitational perturbations of the gas at Lindblad resonances, and the corotation torque. This torque is generated by the gas on orbital periods very close to the planet's, entering into horseshoe orbits around the planet.

There are different migration regimes depending on the planetary mass.

*2.3.1 Type I migration*

For low mass planets, there is an imbalance in the torques, where the corotation generally opposes the Lindblad torque, but for simple temperature and surface density profiles their net effect is to remove angular momentum from the planet, moving it to smaller orbits.

This is the type I migration regime, with a typical time-scale given by (Tanaka, Takeuchi & Ward 2002):

$$\tau_{migI} = c_{migI} \frac{1}{2.7 + 1.1} \left(\frac{c_s}{a\Omega_k}\right)^2 \left(\frac{M_\star}{M_p}\right) \left(\frac{M_\star}{\Sigma_g a^2}\right) \Omega_k^{-1}$$
$$\simeq 950\,c_{migI} \left(\frac{T}{100\,K}\right) \left(\frac{M_\star}{0.1\,M_\odot}\right)^{1/2} \left(\frac{a}{0.1\,AU}\right)^{1/2}$$
$$\times \left(\frac{M_p}{1\,M_\oplus}\right)^{-1} \left(\frac{\Sigma_g}{10^4\,g\,cm^{-2}}\right)^{-1} \text{yr} \quad (17)$$

were $c_s$ is the sound speed, $\Omega_k$ is the Kepler frequency, and $c_{migI}$ is a parameter that delays migration to take into account uncertainties and potential non-linear effects. We explore $c_{migI} = 1$, 3, and 10 (e.g. Ida & Lin 2004; Ida & Lin 2010).

### 2.3.2 Type II migration

In a viscously evolving disc, a high-mass planet induces a tidal torque that causes the opening of a gap in the orbit. This happens when they reach a mass given by (Ida & Lin 2004; Sasaki, Stewart & Ida 2010; Miguel & Ida 2016):

$$M_{\text{gap}} = 40\alpha \left(\frac{h}{a}\right)^2 M_\star \simeq 1 \left(\frac{\alpha}{10^{-3}}\right) \left(\frac{a}{0.1\,\text{AU}}\right) \left(\frac{T}{100\,\text{K}}\right) M_\oplus \quad (18)$$

with $h$ the disc scale height. Once this happens, the planet is confined in the gap by the Lindblad torques and migrates in a type II migration regime that follows the global disc accretion.

We note that this classical type II migration scenario was recently challenged by Kanagawa, Tanaka & Szuszkiewicz (2018), who performed high-resolution simulations and showed that migration of gap-opening planets might be decoupled from the disc evolution and have a slower migration. Nevertheless, since this mechanism is still under discussion, we use the classical formulation in our calculations and leave the exploration of this mechanism for a future publication.

There are two sub-regimes in type II migration: the disc-dominated and the planet-dominated regimes. The first one is when the local disc mass is higher than the mass of the planet and the time-scale corresponds to the local viscous evolution of the disc. In the second case, the planetary mass is the one that dominates compared to the local disc mass, and then the migration is decelerated by the inertia of the planet. The corresponding time-scales are given by (Hasegawa & Ida 2013):

- disc-dominated time-scale:

$$\tau_{\text{migII,d}} \simeq \frac{M_{\text{disc}}(a)}{\dot{M}_{\text{star}}} \simeq \left(\frac{a}{R_{\text{out}}}\right) \tau_{\text{disc}}$$

$$\simeq 10^3 \left(\frac{a}{0.1\,\text{AU}}\right) \left(\frac{R_{\text{out}}}{100\,\text{AU}}\right)^{-1} \left(\frac{\tau_{\text{disc}}}{10^{-6}}\right) \text{yr} \quad (19)$$

with $R_{\text{out}}$ the outer disc radius, taken as 100 AU in all simulations.

- Planet-dominated regime

$$\tau_{\text{migII,d}} \simeq \text{sign}(a - R_m)\frac{M_p}{\dot{M}_{\text{star}}}, \quad (20)$$

where $R_m$ is the radius of the maximum viscous couple (Lynden-Bell & Pringle 1974; Ida & Lin 2004), where the migration reverses it sign and the planets are push outwards away from the star instead of inwards. During the viscous diffusion process $R_m$ moves outwards quickly following $R_m \simeq 10\,e^{\frac{2t}{\tau_{\text{disc}}}}$ (Ida & Lin 2004).

The critical mass for a gap opening (equation 18) depends on the $\alpha$ considered. Nevertheless, because the different migration regimes have similar time-scales, the choice of $\alpha$ does not affect strongly the synthetic population. In the final population (Section 3) we consider $\alpha = 10^{-3}$ in most of the cases, but we also did some calculations with $\alpha = 10^{-4}$, without finding significant differences in the results.

### 2.3.3 Planetary traps

The mechanism behind planet trapping relies on the fact that disc properties are not simple and transitions in the heating mechanism, dust opacity, or turbulent strength can lead to regions of zero net torque on a planet. In this case the change in the dust opacity across the snow line leads to a local flattening of the temperature gradient which strengthens the corotation torque and reverses the net angular momentum transport between the embryo and the disc. To account for the change in dust opacity across the snow line, we used the temperature-dependent opacity derived in Cridland, Pudritz & Alessi (2019). This function is dependent on two calculations: the first used the results of a detailed astrochemical calculation to derive the water ice distribution as a function of radius across the snow line. Next the dust across the snow line is binned into different populations, each with a different mass fraction of water ice. There are 20 members in the population, each with water ice mass fractions that are evenly spaced in log-space. Hence each population resolves a thin radius range within the water snow line, and the extent of the ice mass fraction covers the minimum and maximum ice abundances inward and outward of the snow line respectively. The effective complex spectral indices for each of these members are computed using the methods outlined by Miyake & Nakagawa (1993), and their resulting opacities are computed using the internal opacity calculator in *RADMC3D* (Dullemond 2012).

The change in the dust opacity impacts the efficiency that the viscously heated part of the disc can shed its heat, hence the model described above adds a small modification to the temperature profile in equation (1). The net torque on the planet is computed following Paardekooper, Baruteau & Kley (2011) and Coleman & Nelson (2014), and is shown in Fig. 5.

A similar change in the temperature profile arises due to the shift in dominant heating source (discussed above), outwards of the water snow line. At this heating transition the temperature profile also becomes more shallow, resulting in a strengthened corotation torque and planet trapping. Along with the water snow line, this heating transition produce two-planet trapping points which effectively negates the inward migration of the forming embryo up to a certain mass.

This maximum mass is related to the size of the corotation region around the protoplanet, which grows as the embryo gains mass. Once the embryo grows to a point when the viscous mixing time becomes shorter than the time it takes for gas to pass around one of the lobes of the gas' horseshoe orbit, then the corotation torque is said to have saturated (see Cridland, Pudritz & Alessi 2016, for details). A saturated corotation torque does not contribute to the net torque on the embryo because any angular momentum that is removed from the embryo around one side of the horseshoe orbit is resupplied on the other side. Hence once the corotation torque saturates, migration is dominated by the Lindblad resonance and the embryo returns to an inward migration (as given by equation 17).

We performed these detailed calculations for the following three cases:

- Case 1: $M_\star = 0.05\,M_\odot$ and disc of $10^{-4}\,M_\odot$, corresponding to a low star and disc mass,
- Case II: $M_\star = 0.1\,M_\odot$ and disc of $10^{-3}\,M_\odot$, corresponding to an intermediate case.
- Case III: $M_\star = 0.25\,M_\odot$ and a disc of $10^{-2}\,M_\odot$, corresponding to a high star and disc mass.

In all three cases we use $\alpha = 10^{-3}$. The results of these simulations show that planets are trapped at $a_{\text{snow}}$ and $a_{\text{vis}-\text{irr}}$, depending on their mass (Fig. 4). At this locations – and for embryos in the right mass range – the direction of the migration is reversed and the planet is trapped.

Based on these detailed and expensive calculations, we adopt a simplified model for the population synthesis calculations, where we assume that a migration-trap exists at $a_{snow}$ and in $a_{vis-irr}$ and at a certain mass range taken from these calculations and interpolated depending on the disc mass of each system (see Guilera et al. 2019, for a more detailed calculation on the effect of this torque on planet formation).

### 2.4 Resonance trapping

When two neighbours protoplanets have convergent orbits, they are going to be captured into a mean motion resonance when one of them reaches the inner disc or a planetary trap and their migration is slowed down (Terquem & Papaloizou 2007; Ogihara & Ida 2009; Coleman & Nelson 2014; Cossou et al. 2014; Liu et al. 2015; Coleman & Nelson 2016). This is an important effect to consider because the TRAPPIST-1 planets are observed in resonant orbits, similar to what we observe in the Jovian satellites. We adopt an approximation, where the dynamical perturbation between the two approaching protoplanets is calculated neglecting the perturbation of other distant objects. In this case, the protoplanets are trapped when the distance between them is given by $b_{trap}$ (Ida & Lin 2010; Sasaki et al. 2010; Miguel & Ida 2016):

$$b_{trap} = 0.16 \left(\frac{m_i + m_j}{M_\odot}\right)^{1/6} \left(\frac{\Delta v_{mig}}{v_k}\right)^{-1/4} R_H \quad (21)$$

with $m_i$ and $m_j$ the masses of the converging protoplanets, $\Delta v_{mig}$ the difference in their migration speed and $v_k$ the Keplerian speed. $b_{trap}$ is approximately 5 $R_H$.

### 2.5 Dynamical interaction after gaseous disc depletion

While the gaseous disc depletion time-scale is between $10^6$ and $10^7$ yr, we run the simulations for $10^8$ yr. After the gas dissipates, the planets stop the migration due to the interaction with the gaseous disc (as described in Section 2.3). Nevertheless, mutual gravitational interactions between the planets might lead to an excitation of their eccentricities and potential mergers. We take this effect into account using the same model as in Ida & Lin (2010), that we briefly describe here, but we refer to that paper for more details. According to this model, the planets have initial eccentricities that are chosen randomly from a Rayleigh distribution with mean values of their Hill eccentricities, and are in the range 0.001–0.01. Then, for each pair of planets, we calculate the time-scale that is needed for their orbits to cross due to their mutual secular perturbations ($\tau_{cross}$). The pair of planets with the shortest time-scale suffers an encounter before any other pair that changes the eccentricity and semimajor axis of the planets involved. We then estimate if the new parameters might lead to a subsequent close encounter with another neighbouring planet. If this is the case, we compute new changes in eccentricity and semimajor axis due to this second encounter. For this group of planets we check if they have overlapping orbits, and in that case we assume that they undergo a strong collision. Finally we merge those planets and use the conservation of orbital energy to estimate the semimajor axis of the merged body. Then we repeat the procedure until the $\tau_{cross}$ of the pairs is larger than the age of the system.

### 2.6 Initial parameters for the population synthesis calculations

In this paper we explore the formation of planets around stars with masses between 0.05 and 0.25 $M_\odot$. Based on observations made with Herschel and ALMA (Joergens et al. 2013; Ricci et al. 2014; Liu et al. 2015; Daemgen et al. 2016; Testi et al. 2016; van der Plas et al. 2016), and taking into account potential uncertainties (Hendler et al. 2017), we adopt discs with masses between $10^{-4} \leq M_{disc} \leq 5 \times 10^{-2}$ $M_\odot$ and a dissipation time-scale between $10^6 \leq \tau_{disc} \leq 10^7$ yr. For all discs we adopt as the outer radius $R_{out} = 100$ AU.

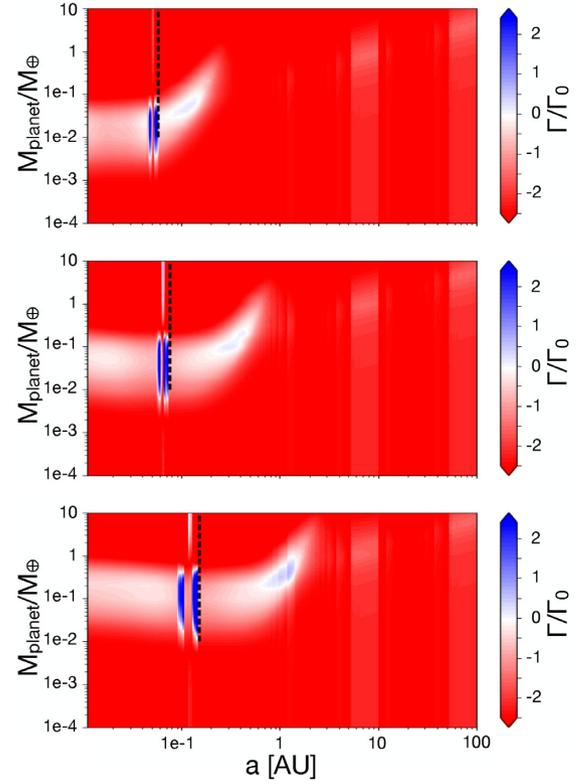

**Figure 5.** Mass versus semimajor axis and torque in different colours, for the three cases considered: low mass star and disc (a), intermediate case (b) and massive star and disc (c) (see the text for details). Migration is inwards in the red areas, it is outwards in the blue areas and the planets are trapped in the white region. The transition between the viscous and irradiated regime is shown with the dashed black line.

For each system, the stellar mass and the disc mass are chosen randomly from linear-uniform distribution with the ranges described above. We check the stability of each system using the Toomre Q parameter (Toomre 1964), defined by:

$$Q \simeq 774.8 \left(\frac{a_{Qmin}}{0.1 AU}\right)^{-3/4} \left(\frac{M_\star}{0.1 M_\odot}\right) \frac{e^{a_{Qmin}/90}}{\Sigma_g^0}, \quad (22)$$

where $a_{Qmin}$ is the semimajor axis for which Q has its minimum value and in this case is $a_{Qmin} = 67.5$ AU (Miguel et al. 2011a). All the systems with $Q > 1$ are stable and are the ones used in our calculations. The gas dissipation time-scale is chosen random from a log-uniform distribution.

Following Ida & Lin (2010) we locate an initial planetary embryo at the inner radius (equation 11) and after that we locate an embryo every $\tilde{\Delta}a$ where:

- $\tilde{\Delta}a$ is the feeding zone corresponding to the local isolation mass, if $a < a_{snow}$, or
- $\tilde{\Delta}a$ is the feeding zone corresponding to the local mass acquired in 1 Gyr, if $a > a_{snow}$

The number of initial embryos in each planetary system varies between many dozens to hundreds according to the disc and stellar mass. The initial size of each planetary embryo is 500 km. An exploration of other values, between 100 and 500 km, showed that the initial size does not affect the final population.

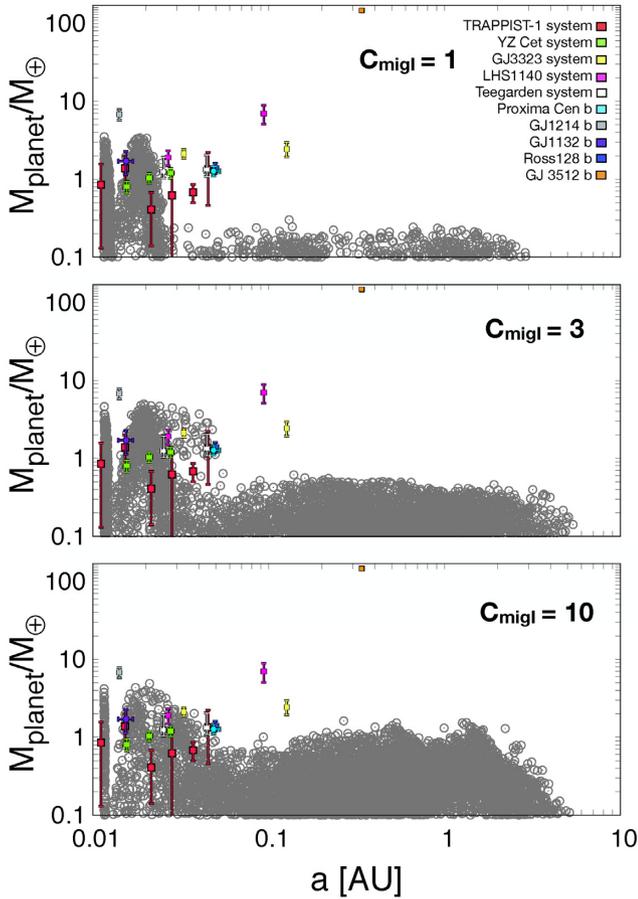

**Figure 6.** Mass versus semimajor axis of the population of synthetic planets formed (grey dots). Observed systems with masses detected with radial velocities are shown as a comparison. Different panels show the populations formed when using different migration scenarios, with $C_{migI}$ = 10 representing a slow migration and $C_{migI}$ = 1 the fastest one. Data for GJ 3512b is from Morales et al. (2019).

## 3 RESULTS

We compute the formation of 3600 planetary systems during $10^8$ yr. We run three sets of simulations with $c_{migI}$ equal to 1, 3, and 10, respectively (Section 2.3), and form 1200 planetary systems in each set. Fig. 6 shows the population of synthetic planets compared with the population of exoplanets – with known masses via precise radial velocities – detected around small stars.

### 3.1 Orbital configuration and masses of the synthetic planets

Our results show that we are able to form most of the exoplanets currently observed around low mass stars with the exception of LHS1140b, GJ1214b, GJ3323c, and GJ 3512b, which have the highest mass and two of them have also the largest semimajor axis observed in the population. We find a bimodal distribution of planets, with planets located at very short periods and masses between 0.1 and up to $\simeq 6\,M_\oplus$ and another subgroup located beyond the snow line ($\simeq 0.1$ AU; equation 5) and masses between 0.1 and up to $\simeq 1.5\,M_\oplus$. The first subgroup are planets that migrate quickly to the inner disc radius ($\simeq 0.01$–0.02 AU; equation 11) – with type I migration if they have small masses, or with type II migration for larger masses ($M_{gap} \simeq 1\,M_\oplus$ for 0.1 AU; equation 18) – and are accumulated there, where they remain until the final stages when the gas dissipates and the collisions and subsequent mergers creates a population of massive planets. The second subgroup is planets that are formed initially beyond the snow line. These planets grow slowly and never reach the mass to open up a gap, and because type I migration rate is slower for larger semimajor axis (equation 17), they do not migrate so efficiently as the ones with shorter periods. In addition, because of the time they take to grow, they only reach large masses when there is no more gas in the disc. Some of these planets might even be trapped in a migration trap if they reach the relevant mass range, but these last ones are the minority.

In general we find that migration traps are not very effective in shaping the population. This is because planets close to the $a_{vis-irr}$ are initially small and migrate quickly to the inner disc before reaching the mass necessary to be trapped in this location. On the other hand, planets close to the snow line might grow fast and acquire the mass necessary to open up a gap and migrate with type II migration. The other possibility for planets beyond the snow line, is that they grow more slowly and do not migrate so efficiently and therefore, when they reach the critical mass to be trapped, there is already very little gas in the disc and they would remain close to the snow line with or without a planetary trap.

Looking at the three different panels, we find that a fast type I migration ($c_{migI}$ = 1) favours the formation of planets with small periods close to the inner disc radius such as the three inner planets in the TRAPPIST-1 system, the two inner planets of YZ Cet and GJ1132b. A scenario with $c_{migI}$ = 3 allows the formation of all planets in the TRAPPIST-1, YZ Cet and Teegarden systems, and also planets such as Proxima Cen b and Ross 128 b and the inner planets in the GJ3323 and LHS1140 systems. Finally, a slower migration rate ($c_{migI}$ = 10) allows the formation of the same observed planets as in the previous case. The main difference between the $c_{migI}$ = 3 and 10 cases is that the last one allows the formation of more massive planets beyond the snow line. Both an intermediate or slow migration rate ($c_{migI}$ = 3 or 10) are able to explain most of the observed exoplanets, with the exception of the most massive ones in the population, probably formed with another, more efficient mechanism.

An interesting analysis comes from the study of the stellar and disc masses that allow the formation of these planetary systems. Fig. 7 shows that only high stellar masses ($M_\star \geq 0.15\,M_\odot$) allow the formation of planets with $M_p \geq 1\,M_\oplus$ and we need relatively high stellar masses ($M_\star \geq 0.07\,M_{sun}$) to form planets with $M_p \geq 0.1\,M_\oplus$, showing that formation of planets around brown dwarfs might be difficult. We note that even though the population of discs and stellar masses are biased towards large masses (because we use a linear-uniform distribution), this does not change the analysis of the results. The total number of discs with $M_{disc} < 0.01\,M_\odot$ is $\simeq 700$ (approximately 20 per cent of the total population), and only bodies with $M_p < 0.1 M_\oplus$ form in such discs, which confirms that large disc masses are needed to form large planets. This is because the time-scale for solid accretion (equation 16) has a strong dependence on the solid surface density, therefore a more massive disc allows the formation of more massive planets. In addition to this, a more massive gaseous disc also implies that the migration time-scale is shorter (equation 17), moving the planets faster to the inner disc radius and not giving them enough time to grow. Therefore, we have a strong correlation with disc masses: we only form planets with $M_p \geq 0.1\,M_\oplus$ when we have discs masses higher than $M_{disc} \geq 10^{-2} M_{sun}$. Similarly, low stellar masses lead to larger formation time-scales and short migration time-scales. As seen from equations (16) and (17):

$$\frac{\tau_{acc}}{\tau_{migI}} \propto M_p^{5/3} M_\star^{-2/3}, \tag{23}$$

which shows that low stellar masses lead to less efficient planet formation and this effect is more pronounced when we go to higher planetary masses. In addition to this, the snow line of more massive stars is located further out in the disc, and since the planets form preferably in the snow line, this provides the planets more solid material to grow while they migrate inwards. This result is in agreement with results found for higher mass stars, that show that either we are underestimating the disc masses or the planet formation is a much more effective mechanism than currently thought (Manara, Morbidelli & Guillot 2018).

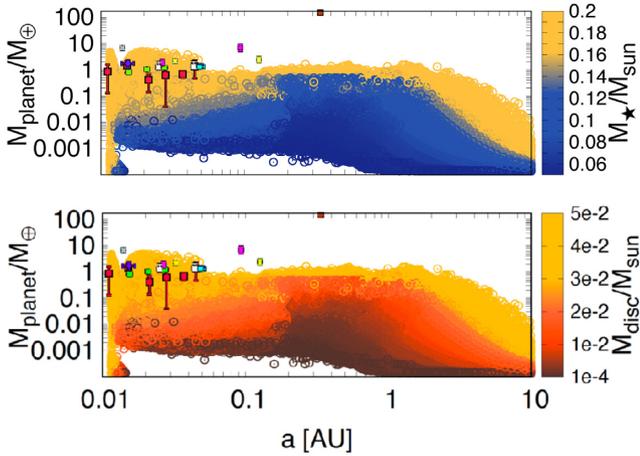

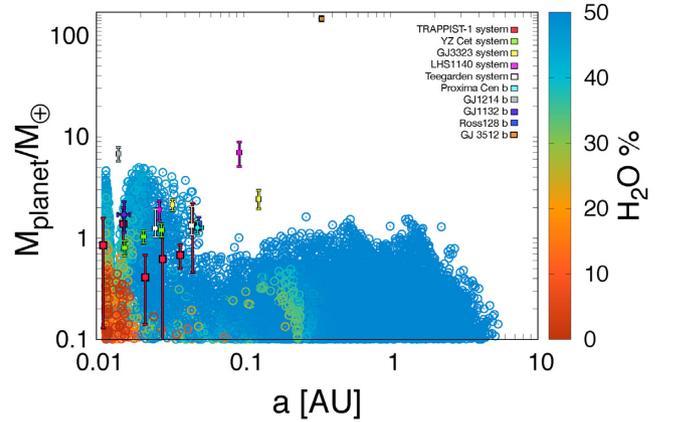

**Figure 7.** Mass and semimajor axis of all the bodies formed in the simulations and their dependence with the stellar (top panel) and disc masses (bottom panel).

**Figure 8.** Mass versus semimajor axis of the synthetic planets formed (all migration scenarios included). The water content is shown by the colour code.

### 3.2 Diversity in planetary composition

In our formation scenario, planets grow first by accreting material from their feeding zones and later due to the collision with neighbouring planets. Consequently, the planets' composition can be directly inferred from the region in the disc where they formed and migrated, and at later stages, with the composition of the planets involved in the mergers. In the inner region of the disc ($a < a_{snow}$) the temperature is high and water is in gaseous phase, thus the solids' composition is mostly rocks. On the other hand, in the outer regions of the disc, ices are the dominant composition for the planetesimals. In order to know the water versus rock content in the formed planets, we adopt a simple model based on Solar system measurements (Abe et al. 2000), that was also used to explore exoplanets (Raymond, Quinn & Lunine 2006; Mandell, Raymond & Sigurdsson 2007; Ronco & de Elia 2014). In this model, the initial planetesimals and planetary embryos have a water content, $w(a)$, that is given by an error function such as it is 0.001 per cent for $a < a_{snow}$, 50 per cent for $a > a_{snow}$, and gradually increases in the intermediate region. We note that the maximum value of 50 per cent is based in Solar system measurements and might change in different planetary systems.

In our model, the planets change their initial composition with time, depending on the ices-to-rocks content of the planetesimals they accrete on their migration paths and, later on, depending on the composition of the planets they merge with if they suffer any collision. The resulting composition of the planets formed is shown in Fig. 8 and a more detailed statistical analysis can be seen in Fig. 9. Our results show that most of the planets with masses $M_p \geq 1 M_\oplus$ have $w(a) > 40$ per cent. These planets were initially located at $a > a_{snow}$, and accreted material with a high water content before they migrated to the inner disc radius. There is a small population of planets ($\simeq 3$ per cent) with $M_p \geq 1 M_\oplus$ and semimajor axis $a \simeq a_{in}$, which have between 30 and 40 per cent of water and are most likely the result of a merge between embryos with different compositions. The population of planets with $M_p < 1 M_\oplus$ presents a larger diversity. These planets have $w(a) \simeq 50$ per cent if they are located beyond the snow line and are mostly made by rocks if they are located close to $a_{in}$, having intermediate values in between. As explained in Section 3.1, since migration time-scale is longer for small planets with large semimajor axis, planets beyond the snow line do not migrate as fast as the ones located in the inner regions of the disc, and they either remain close to the regions where they were originally located or their migration stops once they reach the migration trap at the snow line, having large final water contents during their whole formation.

Our analysis also shows that most of the exoplanets detected around small stars might have a large water content, with the exception of the inner planet in the TRAPPIST-1 system, that might be predominantly made by rocks, in agreement with findings by Unterborn et al. (2018). We note that Dorn et al. (2018) found lower water content for all the TRAPPIST-1 planets, between 0 and 25 per cent. On the other hand, studies made in the habitable zone of M stars have shown that the final water content of the planets might change in their subsequent evolution either due to a runaway phase that can lead to the loss of several Earth oceans of water (Luger & Barnes 2015) or due to high impact speeds that might remove large amounts of volatile material (Lissauer 2007).

When comparing to other planetesimal-driven formation models, our results for planets with $M_p \leq 1 M_\oplus$ are in agreement with the nominal case found by Alibert & Benz (2017), although we are finding a population of more massive planets ($1 < M_p < 5 M_\oplus$) with high water content that is not found in their study. This is probably due to the evolution of planets after the disc dispersal. We find that the most massive planets are the result of mergers between planets of $\simeq 1 M_\oplus$, an scenario not considered by Alibert & Benz (2017). In a recent paper, Coleman et al. (2019) studied the efficiency of planet formation around small stars using N-body simulations, finding similar results to our calculations regarding the maximum planetary mass, the possibility of forming systems with a large number of planets and the large water percentage. In contrast, we are finding a population of planets at larger semimajor axis (that extend beyond the snow line) not found in their calculations, probably due to differences in the migration scenarios (they only consider type I migration without migration trapping). Ogihara & Ida (2009) studied the formation of planets around M-stars, including post-disc-dispersal evolution, with N-body simulations. Their results show that terrestrial planets in the habitable zone of M stars have a high water percentage and their migration speed determines their final orbital configuration, in agreement with our calculations. On the other hand, results by Raymond, Scalo & Meadows (2007) and Ciesla et al. (2015), show mostly dry planets orbiting the habitable zones of low mass stars. While our results for planets with $M_p < 1 M_\oplus$ have more rocks and therefore agree with their calculations, we also find Earth-size (and more massive planets) with large water contents. This difference comes from the fact that they did not include planetary migration, therefore, the planets they formed inside the ice line were mostly rocks, while in our scenario, planetary migration makes planets form beyond the ice line with a large water percentage and then migrate inwards, bringing water-rich planets to regions that were initially dry in the planetary system.

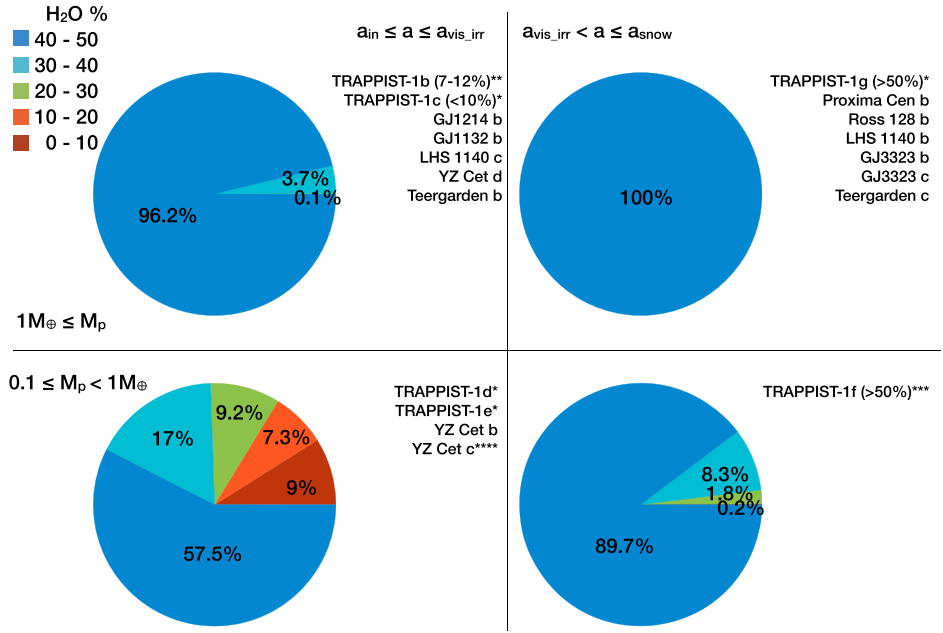

**Figure 9.** Pie charts showing the percentage of planets by their water fractions depending on their mass and semimajor axis. Planets with $M_p \geq 1 M_\oplus$ are in the top panels, those on the left have $a \leq a_{\rm vis-irr}$ and the ones on the right have $a_{\rm vis-irr} < a \leq a_{\rm snow}$. The bottom panels are planets with $0.1 \leq M_p < 1 M_\oplus$ and $a \leq a_{\rm vis-irr}$ (bottom left) and $a_{\rm vis-irr} < a \leq a_{\rm snow}$ (bottom right). Exoplanets are located next to the corresponding pie chart depending on their mass and semimajor axis. *Water fraction for the TRAPPIST-1 planets are from Unterborn et al. (2018). The authors found that any value could be possible for TRAPPIST-1d and TRAPPIST-1e. **TRAPPIST-1b could also be in the category $<1 M_\oplus$ due to errors in the mass determination. ***TRAPPIST-1f can have $>1 M_\oplus$, if we account for the error in mass. ****YZ Cet c has a large error in mass and could have $>1 M_\oplus$.

### 3.3 Architecture of the planetary systems

Because there are few exoplanetary systems discovered around small stars, we have limited knowledge about the most common architectures of such systems. Here we show the trends in the architectures that we find in our simulations. We do not study each system in detail, and instead show general trends, because of the possible uncertainties in the treatment of the resonances and interactions among the planets after the gas dispersal.

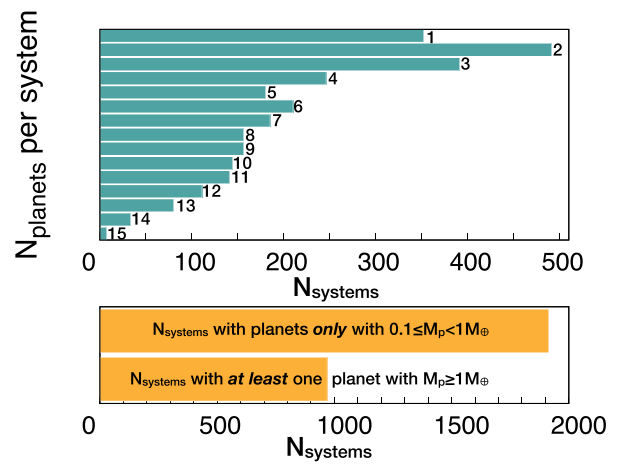

**Figure 10.** Histogram showing the number of planets per system (top panel) for the 2883 systems that have objects with $M_p \geq 0.1 M_\oplus$. The bottom panel shows the number of planetary systems that have only small planets (with masses $0.1 \leq M_p < 1 M_\oplus$) and the ones that have at least one planet with $M_p \geq 1 M_\oplus$.

Fig. 10 shows the number of planets per planetary system, where a 'planet' in this study is an object with $M_p \geq 0.1 M_\oplus$. We see that most of the synthetic planetary systems have 1, 2, or 3 planets. We also see that having 4, 5, and even 6 or 7 planets is still pretty common, and there are a some few systems with many small planets (up to 15). In total, we find 2883 synthetic planetary systems with planets, which represents ∼80 per cent of the total population of 3600 synthetic systems formed. The remaining 20 per cent are systems that formed around low mass discs and stars and are not able to form large bodies (see also Fig. 7). We also analyse the number of planetary systems with small ($0.1 < M_p < 1 M_\oplus$) versus large ($M_p \geq 1 M_\oplus$) planets, and find that approximately one third of the planetary systems that have planets have at least one big planet.

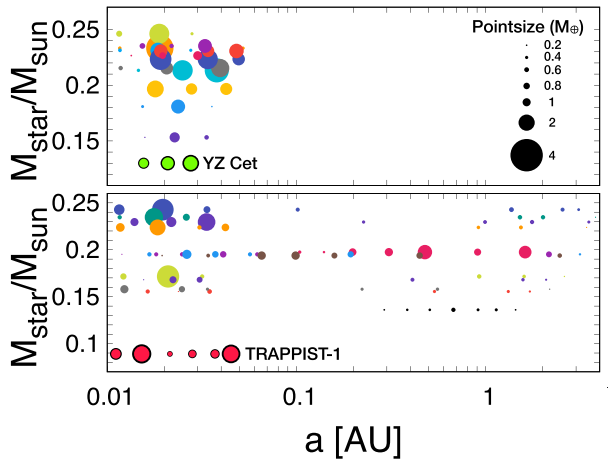

**Figure 11.** Examples of architectures of planetary systems found with three (top panel) and six or seven planets (bottom panel). TRAPPIST-1 and YZ Cet systems are shown as a comparison. Planets are in different size according to their mass.

From the observations, we know that the TRAPPIST-1 system has seven planets located within 0.1 AU, while the YZ Cet system has three planets within 0.03 AU and the GJ3323 and LHS 1140 systems both have two planets: a small one located within 0.04 AU and a more massive one around 0.1 AU. Both of these massive planets at larger periods are difficult to form with the scenario explored in this paper. On the other hand, the recently discovered Teergarden system has two planets that are reproduced by our calculations. In Fig. 11 we show some examples of planetary systems formed in our simulations with three and seven or six planets and compare them with the observed exoplanetary systems. We see that many of the systems we form with three planets have similar characteristics as the YZ Cet system, although the stellar masses are larger. On the other hand, even though it is pretty common to form planetary systems with seven planets, these systems seem to be less compact than the TRAPPIST-1 system, with planets up to a ∼few AU.

## 4 CONCLUSIONS

With the discovery of planetary systems such as the TRAPPIST-1 system, the era of detection and characterization of small exoplanets around M stars and brown dwarfs is starting. Statistical studies show that small, rocky planets should be common around low mass stars (Bonfils et al. 2013; Kopparapu 2013; Dressing & Charbonneau 2015; Hendler et al. 2017), and because these are also the most abundant stars (Henry et al. 2006; Winters et al. 2015), we expect that future surveys (e.g. SPECULOOS project), and space missions (TESS, PLATO) will increase the known population of planets around small nearby stars. In this paper, we explore the properties and diversity of exoplanets around such small stars – e.g. systems similar to the TRAPPIST-1. We form a synthetic population of 3600 planetary systems that evolve $10^8$ yr around different low mass stars and disc environments. To form this synthetic population, we adapt a code that was originally built to study the formation of the Galilean moons, with the addition of migration traps, that were not considered before. In our scheme, the planets form in the core accretion scenario, and they migrate due to the interaction with the gaseous disc that leads to type I and type II migration. Different heating mechanisms and disc properties can lead to a halt in the migration – and trapping of the planets – in regions of zero torque. The solid disc evolves with time due to the planetary accretion and because of the aerodynamic decay, and the gaseous disc dissipates between $10^6$ and $10^7$ yr due to the accretion on to the star. We also consider the subsequent evolution of planets after the gas disc dissipates, with possible orbital crossing and mergers between the planets.

Different to what happens with planets around solar type stars, we find that planets around small stars are mostly rocky and icy because they never reach the critical mass to start gas accretion. In addition, small stars are colder than solar type stars, therefore the snow line is located at small semimajor axis ($\leq 0.2$ AU) and since this is the region where the larger protoplanets emerge, this gives rise to compact systems with planets of short periods that extend up to a few AUs.

As a consequence of planet migration, we find a bimodal distribution of planets with a group located at the inner disc radius ($\simeq 0.01 - 0.02$ AU) and another sub-population that extends beyond the snow line. We also explore different migration rates and find that the population formed with intermediate and slow migration rates ($c_{migI} = 3, 10$) explain most of the planets in the current population of exoplanets orbiting small stars. When looking at the stellar and disc masses, we find that we need stars with masses higher than 0.15 $M_{sun}$ to form massive planets ($M_p \geq 1 M_\oplus$) and stars of at least 0.07 $M_{sun}$ to form planets with masses $M_p \geq 0.1 M_\oplus$, indicating that planets around brown dwarfs might not be common. This is due to the fact that low stellar masses lead to larger growth time-scales and shorter migration time-scales, which makes formation of planets very difficult under those conditions. The trend with the disc mass is even more dramatic: because the solid accretion depends strongly with the disc mass we only form planets when we have discs of at least $10^{-2} M_{sun}$. As a consequence, we cannot form the largest exoplanets in the observed population: LHS1140b, GJ1214b, GJ3323c, and GJ 3512b, and when comparing the observed planetary systems architecture with our results, we observe that also a large disc and stellar mass are necessary to form systems with a large number of planets. This is a strong indication that either we are underestimating the mass of the discs around small stars in the observations or the planet formation process around these stars is much more efficient than studied in this paper. In our planetesimal-driven model, the low efficiency arises because most planets form close to the snow line. Beyond the snow line, the disc surface density is quickly too low to spawn mature planets, even though most of the solid mass is located in these regions. Therefore, massive discs are required to explain the presence of Earth-mass systems. Alternatively, a more efficient planet formation mechanism can be invoked, e.g. as in the pebble-driven scenario for the formation of the TRAPPIST-1 system (Ormel & Liu 2018; Schoonenberg et al. 2019).

We also study the diversity of planets, analysing their ice-to-rock ratio. We find that the large majority (96.2 per cent) of planets with $M_p \geq 1 M_\oplus$ and short periods have a large water content of 40–50 per cent and 100 per cent of the planets with this mass and $a_{\text{vis-irr}} < a \leq a_{\text{snow}}$ have those high water percentages. On the other hand, the population of planets with $0.1 \leq M_p < 1 M_\oplus$ has a larger diversity of ice-to-rock ratios. Those planets with $a \leq a_{\text{vis-irr}}$ are mostly (57.5 per cent) water rich, but we also find that $\simeq 18$ per cent are mostly dry and the rest are in between. For the planets with $a > a_{\text{vis-irr}}$, we find that $\simeq 89.7$ per cent have a water content higher than 40 per cent. We note that these are values after $10^8$ yr and the water content of these planets can change due to the evolution and interaction with their host stars.

We analyse the architecture of the planetary systems (with planets of $M_p \geq 0.1 M_\oplus$) and find that most of the synthetic planetary systems have 1, 2, or 3 planets. We also find that the formation of systems with 4, 5, 6, and even 7 planets is a common outcome in our simulations and one third of the planetary systems with planets have at least one planet with $M_p \geq 1 M_\oplus$.

Our study predicts that compact planetary systems around cool M stars are common. We also identify the key ingredients that differentiate the formation of these planets compared to formation around solar type stars and found that planets around brown dwarfs might be difficult to form. Because we identify key properties expected in the exoplanet population around low mass stars, our analysis will aid in the preparation and interpretation of current and future planetary searches.

## ACKNOWLEDGEMENTS

We acknowledge valuable comments by G. Coleman, R. Burn, and an anonymous referee who contributed to improve our paper. YM and JJF wish to acknowledge support from the Kavli Summer Program in Astrophysics. SI is supported by MEXT KAKENHI 18H05438.

This paper has been typeset from a T<sub>E</sub>X/L<sup>A</sup>T<sub>E</sub>X file prepared by the author.